# Deep-Learning-Based Image Segmentation Integrated with Optical Microscopy for Automatically Searching for Two-Dimensional Materials

Satoru Masubuchi[1, *], Eisuke Watanabe[1], Yuta Seo[1], Shota Okazaki[2], Takao Sasagawa[2], Kenji Watanabe[3], Takashi Taniguchi[1, 3], and Tomoki Machida[1, *]

[1] *Institute of Industrial Science, University of Tokyo, 4-6-1 Komaba, Meguro-ku, Tokyo 153-8505, Japan*

[2] *Laboratory for Materials and Structures, Tokyo Institute of Technology, 4259 Nagatsuta, Midori-ku, Yokohama 226-8503, Japan*

[3] *National Institute for Materials Science, 1-1 Namiki, Tsukuba, Ibaraki 305-0044, Japan*

*Correspondence: mastoru@iis.u-tokyo.ac.jp, tmachida@iis.u-tokyo.ac.jp

**Abstract:**

Deep-learning algorithms enable precise image recognition based on high-dimensional hierarchical image features. Here, we report the development and implementation of a deep-learning-based image segmentation algorithm in an autonomous robotic system to search for two-dimensional (2D) materials. We trained the neural network based on Mask-RCNN on annotated optical microscope images of 2D materials (graphene, hBN, $MoS_2$, and $WTe_2$). The inference algorithm is run on a 1024 × 1024 $px^2$ optical microscope images for 200 ms, enabling the real-time detection of 2D materials. The detection process is robust against changes in the microscopy conditions, such as illumination and color balance, which obviates the parameter-tuning process required for conventional rule-based detection algorithms. Integrating the algorithm with a motorized optical microscope enables the automated searching and cataloging of 2D materials. This development will allow researchers to utilize unlimited amounts of 2D materials simply by exfoliating and running the automated searching process.

**Introduction:**

The recent advances in deep-learning technologies based on neural networks have led to the emergence of high-performance algorithms for interpreting images, such as object detection[1, 2, 3, 4, 5], semantic segmentation[4, 6, 7, 8, 9, 10], instance segmentation[11], and image generation[12]. As neural networks can learn the high-dimensional hierarchical features of objects from large sets of training data[13], deep-learning algorithms can acquire a high generalization ability to recognize images, *i.e.*, they can interpret images that they have not been shown before, which is one of the traits of artificial intelligence[14]. Soon after the success of deep-learning algorithms in general scene recognition challenges[15], attempts at automation began for imaging tasks that are conducted by human experts, such as medical diagnosis[16] and biological image analysis[17, 18]. However, despite significant advances in image recognition algorithms, the implementation of these tools for practical applications remains challenging[18] because of the unique requirements for developing deep-learning algorithms which necessitate the joint development of annotated datasets and software[18, 19].

In the field of two-dimensional (2D) materials[20, 21, 22, 23], the recent advent of autonomous robotic assembly systems has enabled high-throughput searching for exfoliated 2D materials and their subsequent assembly into van der Waals heterostructures[24]. These developments were bolstered by an image recognition algorithm for detecting 2D materials on $SiO_2$/Si substrates[24, 25]; however, current implementations have been developed on the framework of conventional rule-based image processing[26, 27], which uses traditional handcrafted image features such as color contrast, edges, and entropy[24, 25]. Although these algorithms are computationally inexpensive, the detection parameters need to be adjusted by experts, with retuning required when the microscopy conditions change. In contrast, deep-learning algorithms for detecting 2D materials are expected to be robust against changes in optical microscopy conditions, and the development of such an algorithm would provide a generalized 2D materials detector that does not require fine-tuning of the parameters.

In general, deep-learning algorithms for interpreting images are grouped into two categories[28]. Fully convolutional approaches employ an encoder–decoder architecture, such as SegNet[7], U-Net[8], and SharpMask[29]. In contrast, region-based approaches employ feature extraction by a stack of convolutional neural networks (CNNs), such as Mask-RCNN[11], PSP Net[30], and DeepLab[10]. In general, the region-based approaches outperform the fully convolutional approaches for most image segmentation tasks when the networks are trained on a sufficiently large number of annotated datasets[11].

In this work, we implemented and integrated deep-learning algorithms with an automated optical microscope to search for 2D materials on $SiO_2$/Si substrates. The neural network architecture based on Mask-RCNN enabled the detection of exfoliated 2D materials while generating a segmentation mask for each object. Transfer learning from the network trained on the Microsoft common objects in context (COCO) dataset[31] enabled the development of a neural network from a relatively small (~2000 optical microscope images) dataset of 2D materials. Owing to the generalization ability of the neural network, the detection process is robust against changes in the microscopy conditions. These properties could not be realized using conventional rule-based image recognition algorithms.

**Results:**

**System architectures and functionalities:**

A schematic diagram of our deep-learning-assisted optical microscopy system is shown in Fig. 1(a), with photographs shown in Fig. 1(b) and (c). The system comprises three components: (i) an autofocus microscope with a motorized XY scanning stage (Chuo Precision); (ii) a customized software pipeline to capture the optical microscope image, run deep-learning algorithms, display results, and record the results in a database; (iii) a set of trained deep-learning algorithms for detecting 2D materials (graphene, hBN, $MoS_2$, and $WTe_2$). By combining these components, the system can automatically search for 2D materials exfoliated on $SiO_2$/Si substrates (Supplementary Movie 1 and 2). When 2D flakes are detected, their positions and shapes are stored in a database, which can be browsed and utilized to assemble van der Waals heterostructures with a robotic system[24]. The key component developed in this study was the set of trained deep-learning algorithms for detecting 2D materials in the optical microscope images. Algorithm development required three steps, namely, preparation of a large dataset of annotated optical microscope images, training of the deep-learning algorithm on the dataset and deploying the algorithm to run inference on optical microscope images.

The deep-learning model we employed was Mask-RCNN[11] (Fig. 1(a)), which predicts objects, bounding boxes, and segmentation masks in images. When an image is input into the network, the deep convolutional network ResNet101[32] extracts the position-aware high-dimensional features. These features are passed to the region proposal network (RPN) and the region of interest alignment network (ROI Align), which propose candidate regions where the targeted objects are located. The full connection network performs classification (Class) and regression for the bounding box (BBox) of the detected objects. Finally, the convolutional network generates segmentation masks for the objects using the output of the ROI Align layer. This model was developed on the Keras/TensorFlow framework[33, 34, 35].

Before describing dataset preparation and Mask-RCNN training, we show the inference results for optical microscope images containing 2D materials. Fig. 1(c)–(f) show optical microscope images of graphene, $WTe_2$, $MoS_2$, and hBN flakes, which were input into the neural network. The inference results shown in Fig. 1(g)–(j) consist of bounding boxes (colored squares), class labels (text), confidences (numbers), and masks (colored polygons). For the layer thickness classification, we defined three categories: "mono" (1 layer), "few" (2–10 layers), and "thick" (10–40 layers). Note that this categorization was sufficient for practical use in the first screening process because final verification of the layer thickness can be conducted by computational process such as color contrast clustering analysis[25, 36, 37, 38, 39, 40]. As indicated in Fig. 1(g)–(j), the 2D flakes are detected by the Mask-RCNN, and the segmentation mask exhibits good overlap with the 2D flakes. The layer thickness was also correctly classified, with monolayer graphene classified as "mono". The detection process is robust against contaminating objects such as scotch tape residue, particles, and corrugated 2D flakes (white arrows, Fig. 1(e), (f), (i), and (j)).

As the neural network locates 2D crystals using the high-dimensional hierarchical features of the image, the detection results were unchanged when the illumination conditions were varied (Supplementary Movie 3). Fig. 2(a)–(c) shows the deep-learning detection of graphene flakes under differing illumination intensities ($I$). For comparison, the results obtained using conventional rule-based detection are presented in Fig. 2(d)–(f). For the deep-learning case, the results were not affected by changing the illumination intensity from $I = 220$ (a) to 180 (b) or 90 (c) (red, blue, and green curves, Fig. 2). In contrast, with rule-based detection, a slight decrease of the light intensity from $I = 220$ (d) to 200 (e) affected the results significantly, and the graphene flakes became undetectable. Further decreasing the illumination intensity to $I = 180$ (f) resulted in no objects being detected. These results demonstrate the robustness of the deep-learning algorithms over conventional rule-based image processing for detecting 2D flakes.

The deep-learning model was integrated with a motorized optical microscope by developing a customized software pipeline using C++ and Python. We employed a server/client architecture to integrate the deep-learning inference algorithms with the conventional optical microscope (Fig. S1). The image captured by the optical microscope is sent to the inference server, and the inference results are sent back to the client computer. The deep-learning model can run on a graphics-processing unit (NVIDIA Tesla V100) at 200 ms. Including the overheads for capturing images, transferring image data, and displaying inference results, frame rates of ~1 fps were achieved. To investigate the applicability of the deep-learning inference to searching for 2D crystals, we selected $WTe_2$ crystals as a testbed because the exfoliation yields of transition metal dichalcogenides are significantly smaller than graphene flakes. We exfoliated $WTe_2$ crystals onto $1 \times 1$ cm$^2$ $SiO_2$/Si substrates and then conducted searching, which was completed in 1 h using a 50× objective lens. Searching identified ~25 $WTe_2$ flakes on $1 \times 1$ cm$^2$ $SiO_2$/Si with various thicknesses (1–10 layers; Fig. S2).

To quantify the performance of the Mask-RCNN detection process, we manually checked over 2300 optical microscope images, and the detections metrics are summarized in Table S1. Here, we defined true and false positive detections (*TP* and *FP*) as whether the optical microscope image contained at least one correctly detected 2D crystal or not (examples are presented in Fig. S2–Fig. S7). An image in which the 2D crystal was not correctly detected was considered a false negative (*FN*). Based on these definitions, the value of precision was $TP/(FP + FP) \sim 0.53$, which implies that over half of the optical microscope images with positive detection contained $WTe_2$ crystals. Notably, the recall ($TP/(TP + FN) \sim 0.93$) was significantly high. In addition, the examples of false negative detection contain only small fractured $WTe_2$ crystals, which cannot be utilized for assembling van der Waals heterostructures. These results imply that the deep-learning-based detection process does not miss 2D crystals. This property is favorable for the practical application of deep-learning algorithms to searching for 2D crystals, as exfoliated 2D crystals are usually sparsely distributed over $SiO_2$/Si substrates. In this case, false positive detection is less problematic than missing 2D crystals (false negative). The screening of the results can be performed by a human operator without significant intervention[41]. In the case of graphene (Table S1), the precision and recall were both high (~0.95 and ~0.97, respectively), which implies excellent performance of the deep-learning algorithm for detecting 2D crystals. We speculate that there is a difference between the exfoliation yields of graphene and WTe$_2$ because the mean average precision (mAP) at the intersection of union (IOU) over 50% mAP@IoU$_{50\%}$ with respect to the annotated dataset (see preparation methods below) for each material does not differ significantly (0.49 for graphene and 0.52 for WTe$_2$). As demonstrated above, these values are sufficiently high and can be successfully applied to searches for 2D crystals. These results indicate that the deep-learning inference can be practically utilized to search for 2D crystals.

**Preparation of training data:**

To develop the Mask-RCNN model presented above, we prepared annotated images and trained networks as follows. In general, the performance of a deep-learning network is known to scale with the size of the dataset[42]. To collect a large set of optical microscope images containing 2D materials, we exfoliated graphene (covalent material), MoS$_2$ (2D semiconductors), WTe$_2$, and hBN crystals onto SiO$_2$/Si substrates. Using the automated optical microscope, we collected ~2100 optical microscope images containing graphene, MoS$_2$, WTe$_2$, and hBN flakes. The images were annotated manually using a web-based labeling tool[43]. Fig. 3(a) shows representative annotated images, and Fig. 3(b) shows the annotation metrics. The dataset comprises 353 (hBN), 862 (graphene), 569 (MoS$_2$), and 318 (WTe$_2$) images. The numbers of annotated objects were 456 (hBN), 4805 (graphene), 839 (MoS$_2$), and 1053 (WTe$_2$). The annotations were converted to the JSON format compatible with the Microsoft COCO dataset using our customized scripts written in Python. Finally, the annotated dataset was randomly divided into training and test datasets in an 8:2 ratio.

**Model training:**

To train the model on the annotated dataset, we utilized the multitask loss function defined in Ref. 11 and [33]: $L = \alpha L_{cls} + \beta L_{box} + \gamma L_{mask}$, where $L_{cls}$, $L_{box}$, and $L_{mask}$ are the classification, localization, and segmentation mask losses, respectively; $\alpha - \gamma$ is the control parameter for tuning the balance between the loss sets as $(\alpha, \beta, \gamma) = (0.6, 1.0, 1.0)$. The class loss was $L_{cls} = -\log p_u$, where $p = (p_0, \ldots, p_k)$ is the probability distribution for each region of interest in which the result of classification is $u$. The bounding box loss $L_{box}$ is defined as $L_{box}(t^u, v) = \sum_{i \in \{x,y,w,h\}} \text{smooth}_{L_1}(t_i^u - v_i)$, where $\text{smooth}_{L_1}(x) = \begin{cases} 0.5x^2, & |x| < 1 \\ |x| - 0.5, & \text{otherwise} \end{cases}$ is an $L_1$ loss. The mask loss $L_{mask}$ was defined as the average binary cross-entropy loss: $L_{mask} = -\frac{1}{m^2} \sum_{1 \leq i,j \leq m} [y_{ij} \cdot \log \hat{y}_{ij}^k + (1 - y_{ij}) \log(1 - \hat{y}_{ij}^k)]$, where $y_{ij}$ is the binary mask at $(i, j)$ from an ROI of $(m \times m)$ size on the ground truth mask of class $k$, and $\hat{y}_{ij}^k$ is the predicted class label of the same cell.

Instead of training the model from scratch, the model weights, except for the network heads, were initialized using those obtained by pretraining on a large-scale object segmentation dataset in general scenes, *i.e.*, the MS-COCO dataset[31]. The remaining parts of the network weights were initialized using random values. The optimization was conducted using a stochastic gradient decent with a momentum of 0.9 and a weight decay of 0.0001. Each training epoch consisted of 500 iterations. The training comprised four stages, each lasting for 30 epochs (Fig. 3(c)). For the first two training stages, the learning rate was set to $10^{-3}$. The learning rate was decreased to $10^{-4}$ and $10^{-5}$ for the last two stages. In the first stage, only the network heads were trained (top row, Fig. 3(c)). Next, the parts of the backbone starting at layer 4 were optimized (second row, Fig. 3(c)). In the third and fourth stages, the entire model (backbone and heads) was trained (third and fourth rows, Fig. 3(c)). The training took 12 h using four GPUs (NVIDIA Tesla V100 with 32 GB memory). To increase the number of training datasets, we used data augmentation techniques including color channel multiplication, rotation, horizontal/vertical flips, and horizontal/vertical shifts. These operations were applied to the training data with a random probability on-line to reduce disk usage (examples of the augmented data are presented in Fig. S8 and Fig. S9). Before being fed to the Mask-RCNN model, each image was resized to 1024 × 1024 $px^2$ while preserving the aspect ratio, with any remaining space zero padded.

To improve the generalization ability of the network, we organized the training of the Mask-RCNN model into two steps. First, the model was trained on mixed datasets consisting of multiple 2D materials (graphene, hBN, $MoS_2$, and $WTe_2$). At this stage, the model was trained to perform segmentation and classification both on material identity and layer thickness. Then, we use the trained weights as a source and performed transfer learning on each material subset to achieve layer thickness classification. By employing this strategy, the feature values that are common to 2D materials behind the network heads were optimized and shared between the different materials. As shown below, the sharing of the backbone network contributed to faster convergence of the network weights and a smaller test loss.

**Training curve:**

Fig. 3(d) shows the value of the loss function as a function of the epoch count. The solid (dotted) curves represent the test (training) loss. The training was conducted either with (red curves) or without (blue curves) data augmentation. Without augmentation, the training loss decreased to zero, while the test loss was increased. The difference between the test and training losses was significantly increased with training, which indicates that the generalization error increased and the model overfit the training data[13]. When data augmentation was applied, both the training and validation losses decreased monotonically with training, and the difference between the training and validation losses was small. These results indicate that when 2000 optical microscope images are prepared, the Mask-RCNN model can be trained on 2D materials without overfitting.

**Transfer learning from the trained model for accurate deep-learning inference models:**

After training on multiple material categories, we applied transfer learning to the model using each sub-dataset. Fig. 4(a)–(d) show the learning curves for training the networks on the graphene, hBN, $MoS_2$, and $WTe_2$ subsets of the annotated data, respectively. The solid (dotted) curves represent the test (training) loss. The network weights were initialized using those at epoch 120 obtained by training on multiple material classes Fig. 3(d) (red curves, Fig. 4 (a)–(d)). For reference, we also trained the dataset by initializing the network weights using those obtained by pretraining only on the MS-COCO dataset (blue curves, Fig. 4(a)–(d)). Notably, in all cases, the test loss decreased faster for those pretrained on the 2D crystals and MS-COCO than for those pretrained on MS-COCO only. The loss value after 30 epochs of training on 2D crystals and MS-COCO was of almost the same order as that obtained after 80 epochs of training on MS-COCO only. In addition, the minimum loss value achieved in the case of pretraining on 2D crystals and MS-COCO was smaller than that achieved with MS-COCO only. These results indicate that the feature values that are common to 2D materials are learnt in the backbone network. In particular, the trained backbone network weights contribute to improving the model performance on each material.

To investigate the improvement of the model accuracy, we compared the inference results for the optical microscope images using the network weights from each training set. Fig. 4(e) and (h) show the optical microscope images of graphene and $WTe_2$, respectively, input into the network. We employed the model weights where the loss value was minimum (indicated by the red/blue arrows). The inference results in the cases of transferring only from MS-COCO and from both MS-COCO and 2D materials are shown in Fig. 4(f) and (g) for graphene and Fig. 4(i) and (j) for WTe$_2$. For graphene, the model transferred from MS-COCO only failed in detecting some thick graphite flakes, as indicated by the white arrows in Fig. 4(f), whereas the model trained on MS-COCO and 2D crystals detected the graphene flakes, as indicated by the white arrows in Fig. 4(g). Similarly, for WTe$_2$, when the inference process was performed using the model transferred from MS-COCO only, the surface of the SiO$_2$/Si substrate surrounded by thick WTe$_2$ crystals was misclassified as WTe$_2$, as indicated by the white arrow in Fig. 4(d). In contrast, when learning was transferred from the model pretrained on MS-COCO and 2D materials (red arrow, Fig. 4(b)), this region was not recognized as WTe$_2$. These results indicate that pretraining on multiple material classes contributes to improving model accuracy because the common properties of 2D crystals are learnt in the backbone network. The inference results presented in Fig. 1 were obtained by utilizing the model weights at epoch 120 for each material.

**Generalization ability:**

Finally, we investigated the generalization ability of the neural network for detecting graphene flakes in images obtained using different optical microscope setups (Asahikogaku AZ10-T/E, Keyence VHX-900, and Keyence VHX-5000 as shown in Fig. 5(a)–(c), respectively). Fig. 5(d)–(f) show the optical microscope images of exfoliated graphene captured by each instrument. Across these instruments, there are significant variations in the white balance, magnification, resolution, illumination intensity, and illumination inhomogeneity (Fig. 5(d)–(f)). The model weights from training epoch 120 on the graphene dataset were employed (red arrow, Fig. 4(d)). Even though no optical microscope images recorded by these instruments were utilized for training, as shown by the inference results in Fig. 5(g)–(i), the deep-learning model successfully detected the regions of exfoliated graphene. These results indicate that our trained neural network captured the latent general features of graphene flakes and thus constitutes a general-purpose graphene detector that works irrespective of the optical microscope setup. These properties cannot be realized by utilizing the conventional rule-based detection algorithms for 2D crystals, where the detection parameters must be re-tuned when the optical conditions were altered.

**Discussion and summary:**

In order to train the neural network for the 2D crystals which have different appearance such as $ZrSe_3$, the model weights trained on both MS-COCO and 2D crystals obtained in this study can be used as source weights to start training. In our experience, the Mask-RCNN trained on a small dataset of ~80 images from the MS-COCO pretrained model can produce rough segmentation masks on graphene. Therefore, providing less than 80 annotated images would be sufficient for developing a classification algorithm that works for detecting other 2D materials when we use our trained weights as a source. Our work can be utilized as a starting point for developing neural network models which works for various 2D materials.

Moreover, the trained neural networks can be utilized for searching the materials other than those used for training. For demonstration, we exfoliated $WSe_2$ and $MoSe_2$ flakes on $SiO_2/Si$ substrate and conducted searching with the model trained on $WTe_2$. As shown in Fig. S10 and Fig. S11 in supplementary information, thin $WSe_2$ and $MoSe_2$ flakes are correctly detected even without training on these materials. This result indicates that the difference of the appearances of $WSe_2$ and $MoSe_2$ from $WTe_2$ are covered by the generalization ability of neural networks.

Finally, our deep-learning inference process can run on the remote server/client architecture. This architecture is suitable for researchers with an occasional need for deep learning, as it provides a cloud-based setup that does not require a local GPU. The conventional optical microscope instruments that were not covered in this study can also be modified to support deep learning inference by implementing the client software to capture image, send image to the server, receive and display inference results. The distribution of the deep-learning inference system will benefit the research community by saving the time needed for optical-microscopy-based searching of 2D materials.

In this work, we developed a deep-learning-assisted automated optical microscope to search for 2D crystals on $SiO_2$/Si substrates. A neural network with Mask-RCNN architecture trained on 2D materials enabled the efficient detection of various exfoliated 2D crystals, including graphene, hBN, and transition metal dichalcogenides ($WTe_2$ and $MoS_2$), while simultaneously generating a segmentation mask for each object. This work, which should free researchers from the repetitive tasks of optical microscopy, comprises a fundamental step toward realizing fully automated fabrication systems for van der Waals heterostructures.

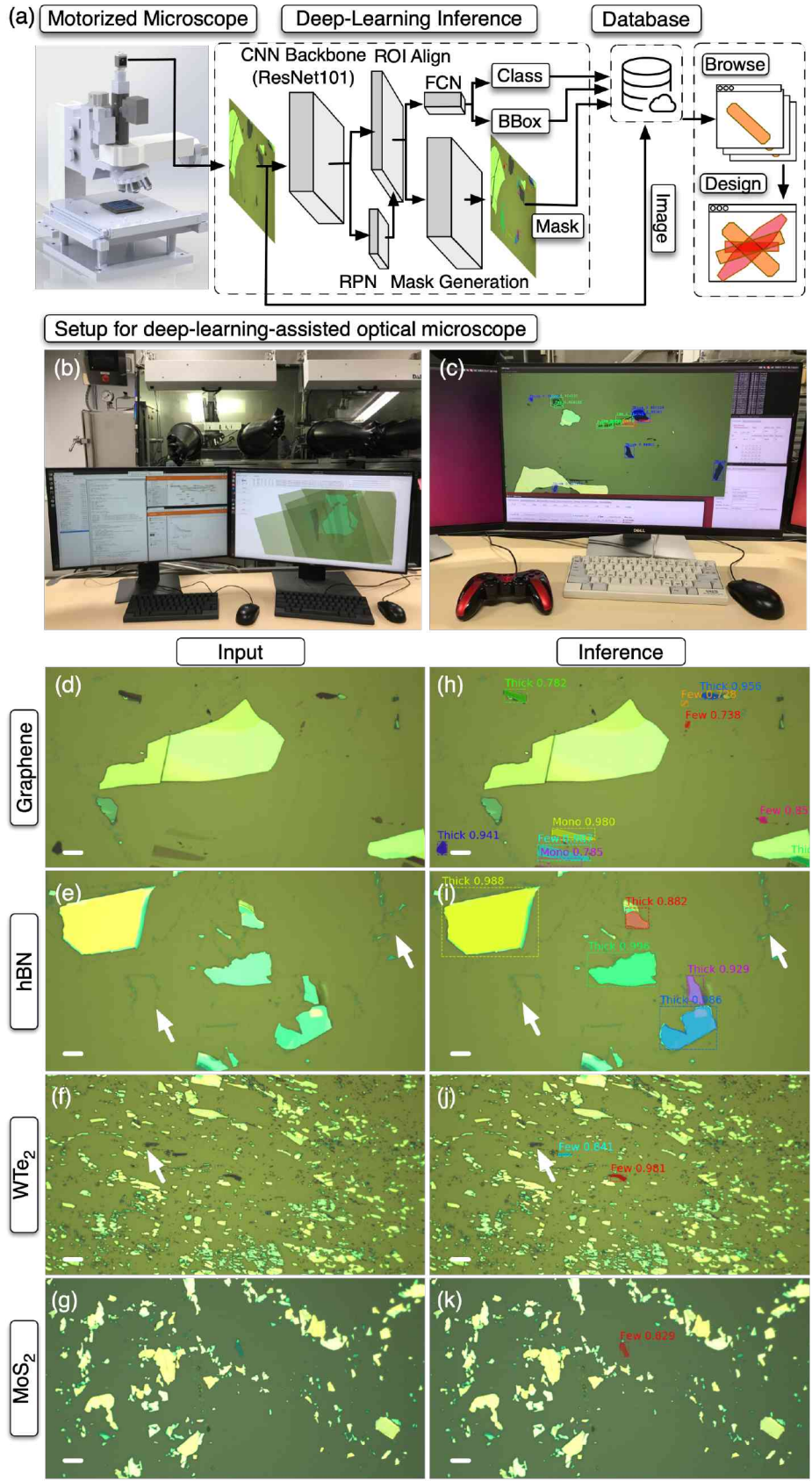

Fig. 1. Deep-learning-assisted automated optical microscope for searching for two-dimensional

(2D) crystals. (a) Schematic of the deep-learning-assisted optical microscope system. The optical microscope acquires an image of exfoliated 2D crystals on a $SiO_2$/Si substrate. The images are input into the deep-learning inference algorithm. The Mask-RCNN architecture generates a segmentation mask, bounding boxes, and class labels. The inference data and images are stored in a cloud database, which forms a searchable database. The customized computer-assisted-design (CAD) software enables browsing of 2D crystals and designing of van der Waals heterostructures. (b) and (c) Photographs of (b) the optical microscope and (c) the computer screen for deep-learning-assisted automated searching. (d)–(k) Segmentation of 2D crystals. Optical microscope images of (d) graphene, (e) hBN, (f) $WTe_2$, and (g) $MoS_2$ on $SiO_2$ (290 nm)/Si. (h)–(k) Inference results for the optical microscope images in (d)–(g), respectively. The segmentation masks and bounding boxes are indicated by polygons and dashed squares, respectively. In addition, the class labels and confidences are displayed. The scale bars correspond to 10 μm.

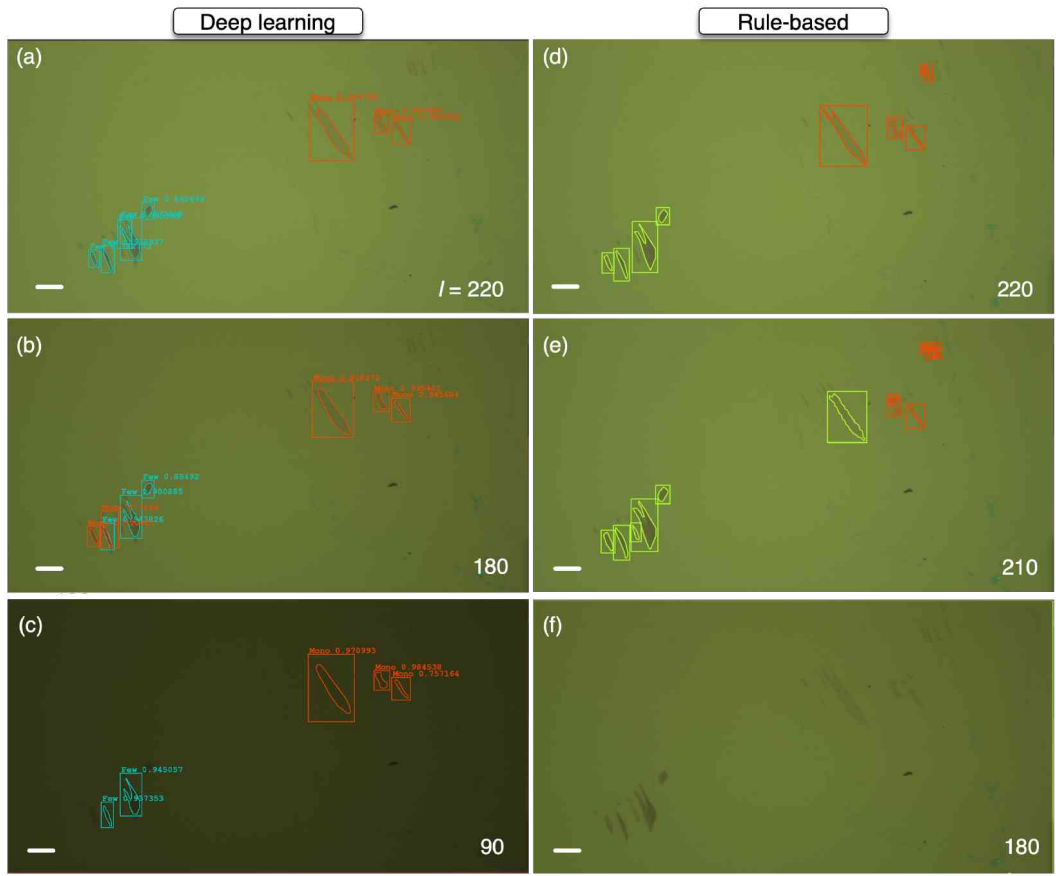

Fig. 2. Comparison between deep-learning and rule-based detection. Input image and inference results under illumination intensities of $I$ = (a) 220, (b) 180, and (c) 90 (arb. unit) for deep-learning detection and $I$ = (d) 220, (e) 210, and (f) 180 (arb. unit) for rule-based detection. The scale bars correspond to 10 μm.

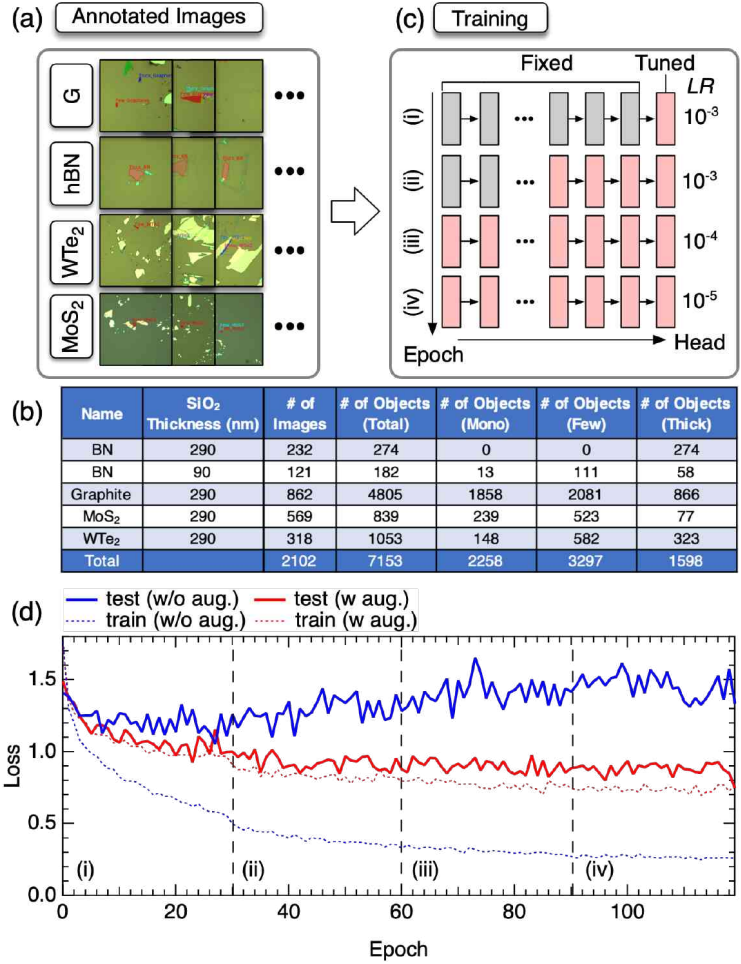

Fig. 3. Overview of training. (a) Examples of annotated datasets for graphene (G), hBN, WTe$_2$, and MoS$_2$. (b) Training data metrics. (c) Schematic of the training procedure. (d) Learning curves for training on the dataset. The network weights were initialized by the model weights pretrained on the MS-COCO dataset. Solid (dotted) curves are test (train) losses. Training was performed either with (red curve) or without (blue curve) augmentation.

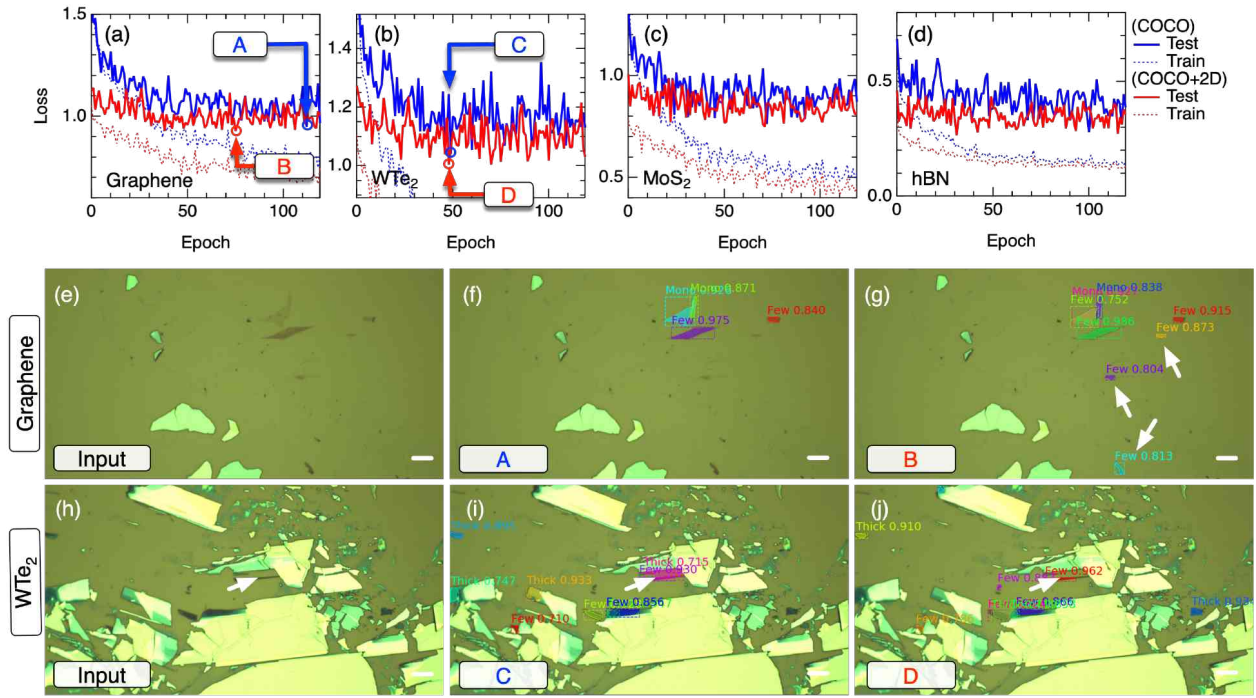

Fig. 4. Transfer learning from model weights pretrained on MS-COCO and MS-COCO+2D datasets. (a)–(d) Test (solid curves) and training (dotted curves) losses as a function of epoch count for training on (a) graphene, (b) $WTe_2$, (c) $MoS_2$, and (d) hBN. Each epoch consists of 500 training steps. The model weights were initialized using those pretrained on (blue) MS-COCO and (red) MS-COCO and 2D materials datasets. The optical microscope image of graphene ($WTe_2$) and the inference results for these images are shown in (e)–(g) ((h)–(j)). The scale bars correspond to 10 μm.

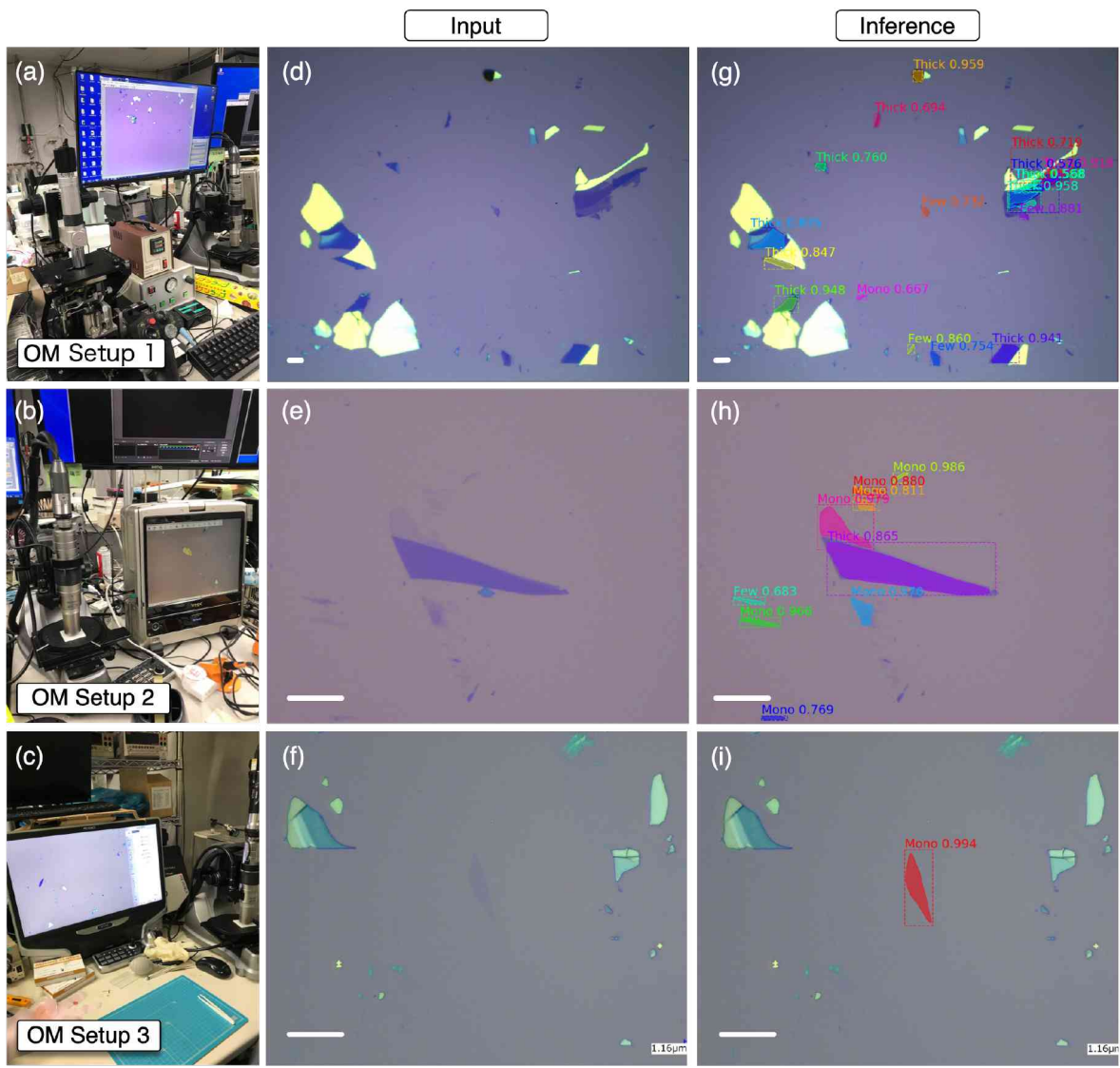

Fig. 5. Generalization ability of the neural network. (a)–(c) Optical microscope setups used for capturing images of exfoliated graphene (Asahikogaku AZ10-T/E, Keyence VHX-900, and Keyence VHX-5000, respectively). (d)–(f) Optical microscope images recorded using instruments (a)–(c), respectively. (g)–(i) Inference results for the optical microscope images in (d)–(f), respectively. The segmentation masks are shown in color, and the category and confidences are also indicated. The scale bars correspond to 10 μm.

## Methods:

### Optical microscope drivers:

The automated optical microscope drivers were written in C++ and Python. The software stack was developed on the stacks of a robotic operating system[44] and the HALCON image processing library (MVTec Software GmbH).

### Preparation of training dataset:

To obtain the Mask-RCNN model to segment 2D crystals, we employed a semi-automatic annotation workflow. First, we trained the Mask-RCNN with a small dataset consisting of ~80 images of graphene. Then, we conducted predictions on optical microscope images of graphene. The prediction labels generated using the Mask-RCNN were stored in LabelBox using API. These labels were manually corrected by a human annotator. This procedure greatly enhanced the annotation efficiency, allowing each image to be labeled in 20–30 s.

### Data availability:

The data that support the findings of this study are available from the corresponding author upon reasonable request.

### Acknowledgements:

This work was supported by the Core Research for Evolutional Science and Technology, Japan Science and Technology Agency (JST), under Grant Nos. JPMJCR15F3 and JPMJCR16F2, and by JSPS KAKENHI under Grant No. JP19H01820.

### Author contributions:

S. M. conceived the scheme, implemented the software, trained the neural network, and wrote the manuscript. E. W. and Y. S. exfoliated the 2D materials and tested the system. S. O. and T. S. synthesized the WTe$_2$ and WSe$_2$ crystals. K. W. and T. T. synthesized the hBN crystals. T. M. supervised the research program.

**Competing Interests:** The authors declare there are no competing interests.

**References:**

1. Zhao Z-Q, Zheng P, Xu S-t, Wu X. Object detection with deep learning: A review. *IEEE Transactions on Neural Networks and Learning Systems* 2019.

2. Ren S, He K, Girshick R, Sun J. Faster R-CNN: Towards real-time object detection with region proposal networks. *Advances in Neural Information Processing Systems* 2015**:** 91-99.

3. Girshick R. Fast R-CNN. *Proceedings of the IEEE International Conference on Computer Vision* 2015**:** 1440-1448.

4. Girshick R, Donahue J, Darrell T, Malik J. Rich feature hierarchies for accurate object detection and semantic segmentation. *Proceedings of the IEEE Conference on Computer Vision and Pattern Recognition* 2014**:** 580-587.

5. Liu W, Anguelov D, Erhan D, Szegedy C, Reed S, Fu C-Y*, et al.* SSD: Single shot multibox detector. *European Conference on Computer Vision* 2016**:** 21-37.

6. Garcia-Garcia A, Orts-Escolano S, Oprea SO, Villena-Martinez V, Garcia-Rodriguez J. A review on deep learning techniques applied to semantic segmentation. *arXiv:170406857* 2017.

7. Badrinarayanan V, Kendall A, Cipolla R. SegNet: A deep convolutional encoder-decoder architecture for image segmentation. *IEEE Transactions on Pattern Analysis and Machine Intelligence* 2017, **39**(12)**:** 2481-2495.

8. Ronneberger O, Fischer P, Brox T. U-Net: Convolutional networks for biomedical image segmentation. *International Conference on Medical Image Computing and Computer-assisted Intervention* 2015**:** 234-241.

9. Long J, Shelhamer E, Darrell T. Fully convolutional networks for semantic segmentation. *Proceedings of the IEEE Conference on Computer Vision and Pattern Recognition* 2015**:** 3431-3440.

10. Chen L-C, Papandreou G, Kokkinos I, Murphy K, Yuille AL. DeepLab: Semantic image segmentation with deep convolutional nets, atrous convolution, and fully connected CRFs. *IEEE Transactions on Pattern Analysis and Machine Intelligence* 2017, **40**(4)**:** 834-848.

11. He K, Gkioxari G, Dollár P, Girshick R. Mask R-CNN. *Proceedings of the IEEE International Conference on Computer Vision* 2017**:** 2961-2969.

12. Goodfellow I, Pouget-Abadie J, Mirza M, Xu B, Warde-Farley D, Ozair S*, et al.* Generative adversarial nets. *Advances in Neural Information Processing Systems* 2014**:** 2672-2680.

13. Goodfellow I, Bengio Y, Courville A. *Deep Learning*. MIT Press, 2016.

14. LeCun Y, Bengio Y, Hinton G. Deep learning. *Nature* 2015, **521:** 436-444.

15. Krizhevsky A, Sutskever I, Hinton GE. Imagenet classification with deep convolutional neural networks. *Advances in Neural Information Processing Systems* 2012**:** 1097-1105.

16. Litjens G, Kooi T, Bejnordi BE, Setio AAA, Ciompi F, Ghafoorian M*, et al.* A survey on deep learning in medical image analysis. *Medical Image Analysis* 2017, **42:** 60-88.

17. Falk T, Mai D, Bensch R, Cicek O, Abdulkadir A, Marrakchi Y*, et al.* U-Net: Deep learning for cell counting, detection, and morphometry. *Nature Methods* 2019, **16**(1)**:** 67-70.

18. Moen E, Bannon D, Kudo T, Graf W, Covert M, Van Valen D. Deep learning for cellular image analysis. *Nature Methods* 2019**:** 1.

19. Karpathy. A. Software 2.0.  2017.  [cited]Available from: https://medium.com/@karpathy/software-2-0-a64152b37c35

20. Novoselov KS, Mishchenko A, Carvalho A, Castro Neto AH. 2D materials and van der Waals heterostructures. *Science* 2016, **353**(6298)**:** aac9439.

21. Novoselov KS, Jiang D, Schedin F, Booth TJ, Khotkevich VV, Morozov SV*, et al.* Two-dimensional atomic crystals. *Proceedings of the National Academy of Sciences of the United States of America* 2005, **102**(30)**:** 10451-10453.

22. Novoselov KS, Geim AK, Morozov SV, Jiang D, Zhang Y, Dubonos SV*, et al.* Electric field effect in atomically thin carbon films. *Science* 2004, **306**(5696)**:** 666-669.

23. Zhang Y, Tan YW, Stormer HL, Kim P. Experimental observation of the quantum Hall effect and Berry's phase in graphene. *Nature* 2005, **438**(7065)**:** 201-204.

24. Masubuchi S, Morimoto M, Morikawa S, Onodera M, Asakawa Y, Watanabe K*, et al.* Autonomous robotic searching and assembly of two-dimensional crystals to build van der Waals superlattices. *Nature Communications* 2018, **9**(1)**:** 1413.

25. Masubuchi S, Machida T. Classifying optical microscope images of exfoliated graphene flakes by data-driven machine learning. *npj 2D Materials and Applications* 2019, **3**(1)**:** 4.

26. Nixon MS, Aguado AS. *Feature Extraction & Image Processing for Computer Vision*. Academic Press, 2012.

27. Szeliski R. *Computer Vision: Algorithms and Applications*. Springer Science & Business Media, 2010.

28. Ghosh S, Das N, Das I, Maulik U. Understanding deep learning techniques for image segmentation. 2019**:** arXiv:1907.06119v06111.

29. Pinheiro PO, Lin T-Y, Collobert R, Dollár P. Learning to refine object segments. *European Conference on Computer Vision* 2016**:** 75-91.

30. Zhao H, Shi J, Qi X, Wang X, Jia J. Pyramid scene parsing network. *Proceedings of the IEEE Conference on Computer Vision and Pattern Recognition* 2017**:** 2881-2890.

31. Lin T-Y, Maire M, Belongie S, Hays J, Perona P, Ramanan D*, et al.* Microsoft COCO: Common objects in context. *European Conference on Computer Vision* 2014**:** 740-755.

32. He K, Zhang X, Ren S, Sun J. Deep residual learning for image recognition. *Proceedings of the IEEE Conference on Computer Vision and Pattern Recognition* 2016**:** 770-778.

33. Abdulla W. Mask R-CNN for object detection and instance segmentation on Keras and TensorFlow. 2017 [cited]Available from: https://github.com/matterport/Mask_RCNN

34. Chollet F. Keras: Deep learning for humans. 2015 [cited]Available from: https://github.com/keras-team/keras

35. Abadi M, Barham P, Chen J, Chen Z, Davis A, Dean J*, et al.* Tensorflow: A system for large-scale machine learning. *12th USENIX Symposium on Operating Systems Design and Implementation* 2016**:** 265-283.

36. Lin X, Si Z, Fu W, Yang J, Guo S, Cao Y*, et al.* Intelligent identification of two-dimensional nanostructures by machine-learning optical microscopy. *Nano Research* 2018, **11:** 6316-6324.

37. Li H, Wu J, Huang X, Lu G, Yang J, Lu X*, et al.* Rapid and reliable thickness identification of two-dimensional nanosheets using optical microscopy. *ACS Nano* 2013, **7**(11)**:** 10344-10353.

38. Ni ZH, Wang HM, Kasim J, Fan HM, Yu T, Wu YH*, et al.* Graphene thickness determination using reflection and contrast spectroscopy. *Nano Letters* 2007, **7**(9)**:** 2758-2763.

39. Nolen CM, Denina G, Teweldebrhan D, Bhanu B, Balandin AA. High-throughput large-area automated identification and quality control of graphene and few-layer graphene films. *ACS Nano* 2011, **5**(2)**:** 914-922.

40. Taghavi NS, Gant P, Huang P, Niehues I, Schmidt R, de Vasconcellos SM*, et al.* Thickness determination of MoS 2, MoSe 2, WS 2 and WSe 2 on transparent stamps used for deterministic transfer of 2D materials. *Nano Research* 2019, **12**(7)**:** 1691-1695.

41. Zhang P, Zhong Y, Deng Y, Tang X, Li X. A survey on deep learning of small sample in biomedical image analysis. *arXiv:190800473* 2019.

42. Hestness J, Narang S, Ardalani N, Diamos G, Jun H, Kianinejad H*, et al.* Deep learning scaling is predictable, empirically. *arXiv:171200409* 2017.

43. LabelBox.   [cited]Available from: https://labelbox.com

44. Quigley M, Conley K, Gerkey B, Faust J, Foote T, Leibs J*, et al.* ROS: an open-source Robot Operating System. *ICRA Workshop on Open Source Software* 2009.

**(Supplementary Information)**

**Deep-Learning-Based Image Segmentation Integrated with Optical Microscopy for Automatically Searching for Two-Dimensional Materials**

Satoru Masubuchi[1,*], Eisuke Watanabe[1], Yuta Seo[1], Shota Okazaki[2], Takao Sasagawa[2], Kenji Watanabe[3], Takashi Taniguchi[1,3], and Tomoki Machida[1,*]

[1] *Institute of Industrial Science, University of Tokyo, 4-6-1 Komaba, Meguro-ku, Tokyo 153-8505, Japan*

[2] *Laboratory for Materials and Structures, Tokyo Institute of Technology, 4259 Nagatsuta, Midori-ku, Yokohama 226-8503, Japan*

[3] *National Institute for Materials Science, 1-1 Namiki, Tsukuba, Ibaraki 305-0044, Japan*

*Correspondence: mastoru@iis.u-tokyo.ac.jp, tmachida@iis.u-tokyo.ac.jp

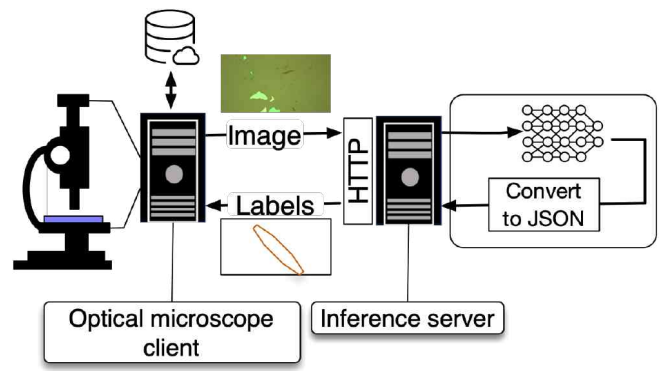

Fig. S1. Server/client architecture employed to integrate the deep-learning algorithms with the automated optical microscope to search for 2D flakes.

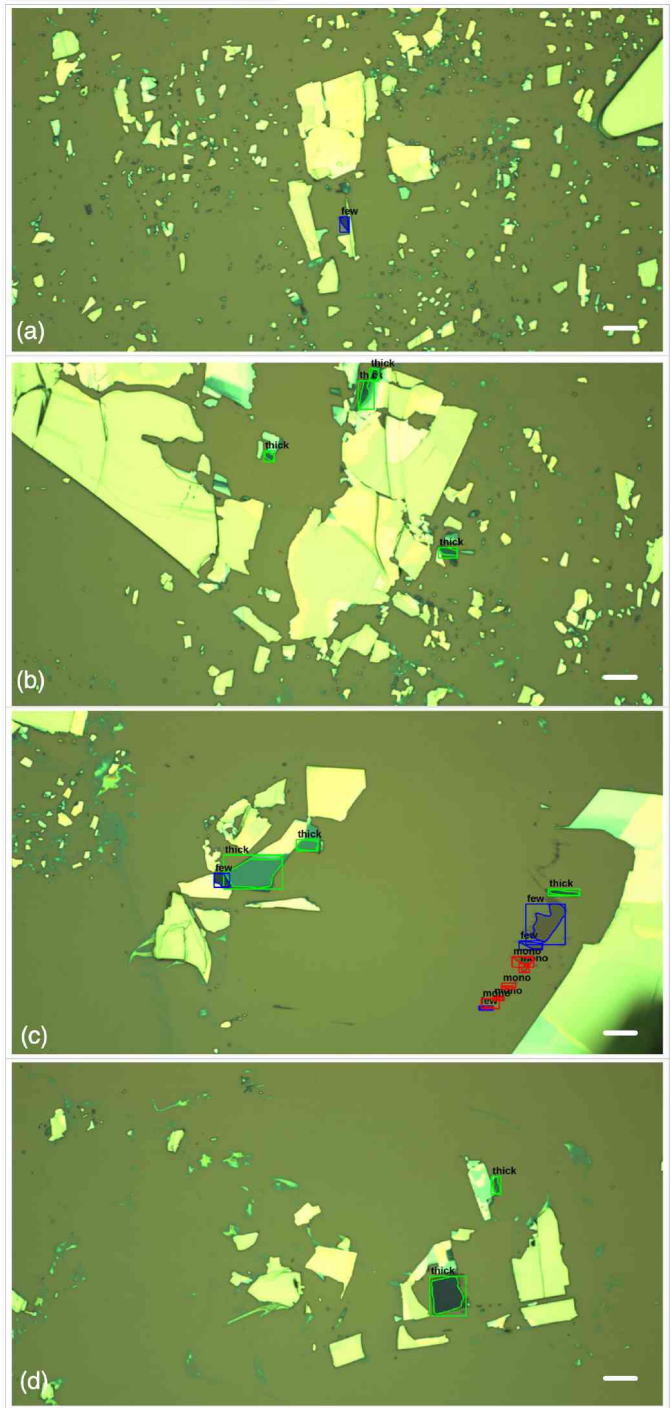

Fig. S2. Examples of true positive detection for WTe$_2$ flakes. The scale bars correspond to 10 µm.

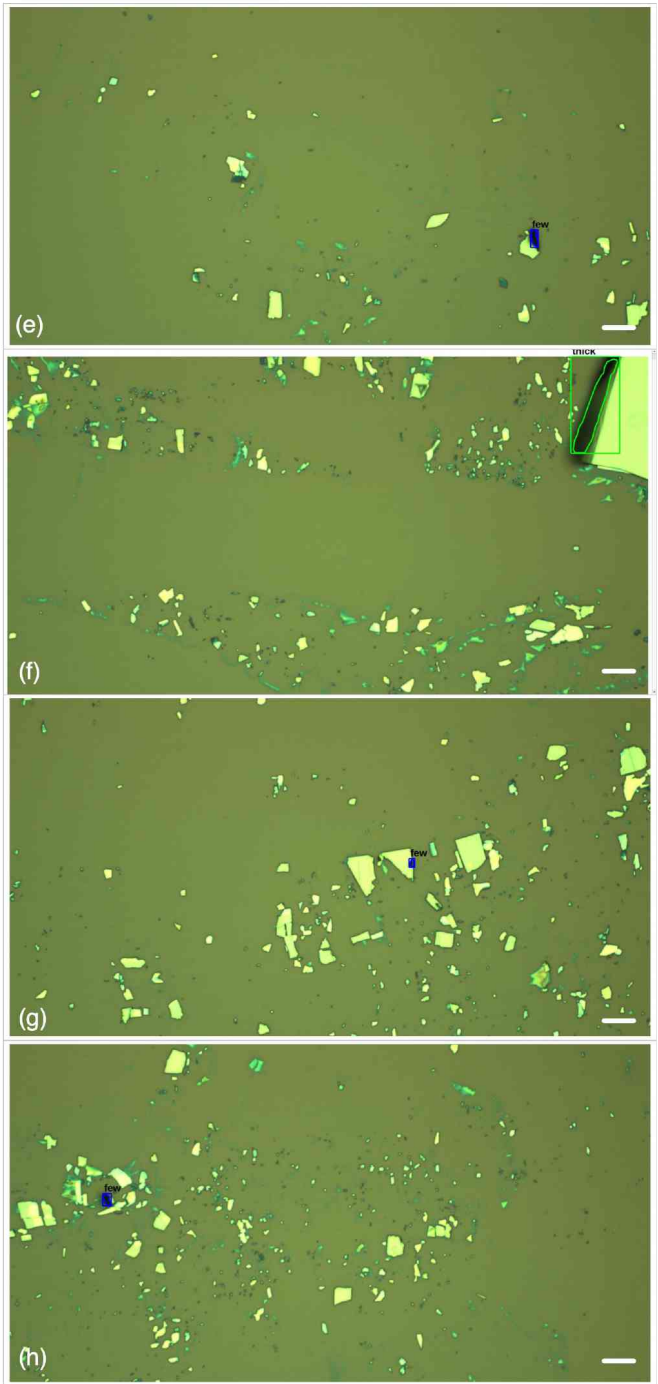

Fig. S3. Examples of false positive detection for WTe$_2$ flakes. The scale bars correspond to 10 µm.

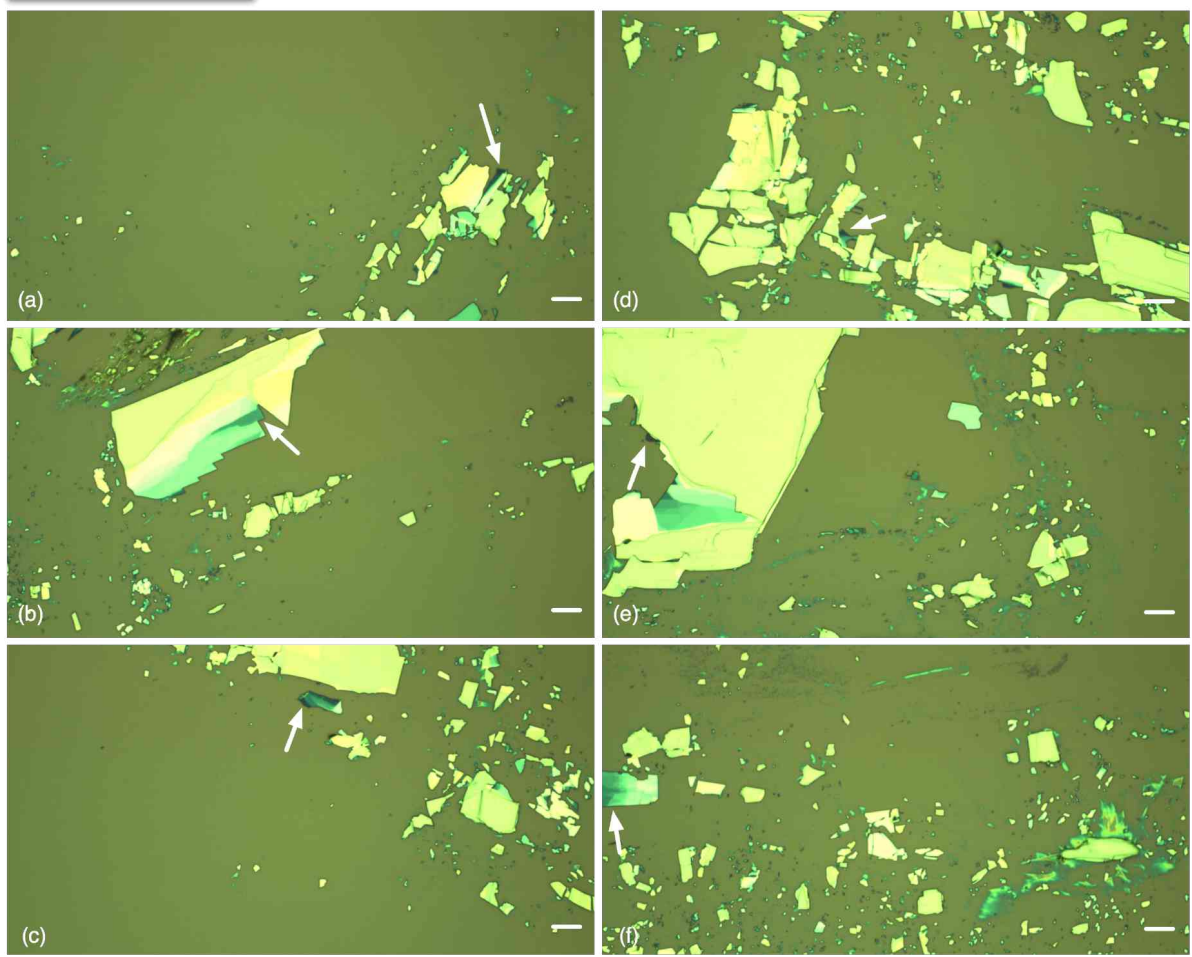

Fig. S4. Examples of false negative detection for WTe$_2$ flakes. The positions of the WTe$_2$ flakes are indicated by the white arrows. The scale bars correspond to 10 μm.

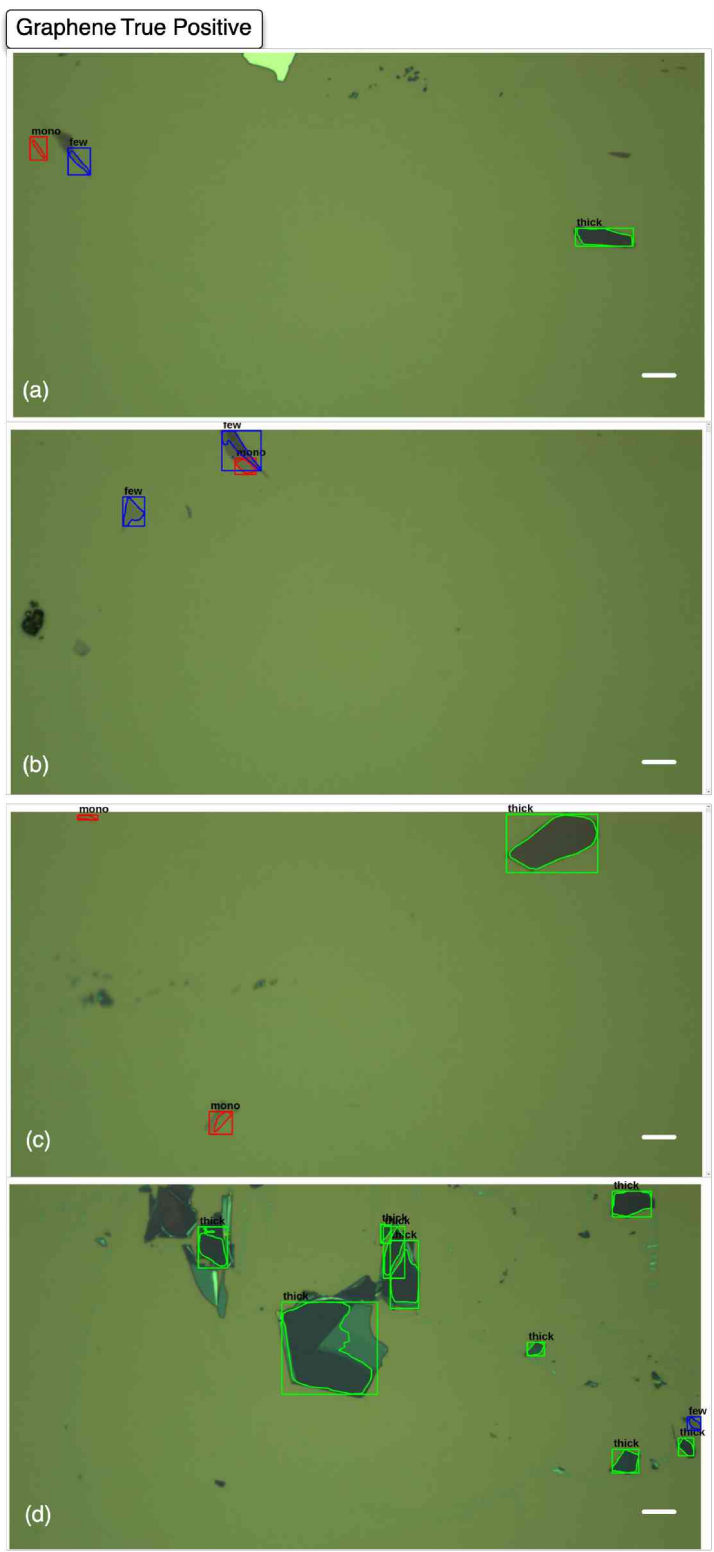

Fig. S5. Examples of true positive detection for graphene. The scale bars correspond to 10 µm.

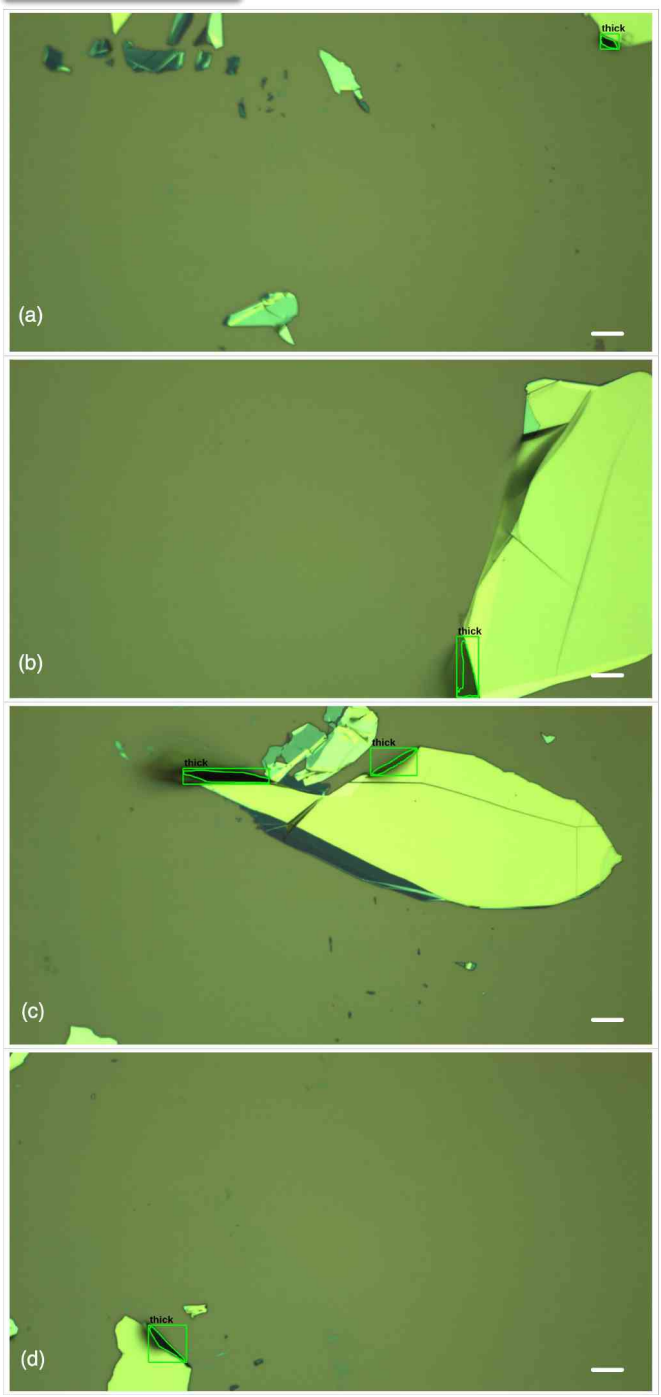

Fig. S6. Examples of false positive detection for graphene. The scale bars correspond to 10 µm.

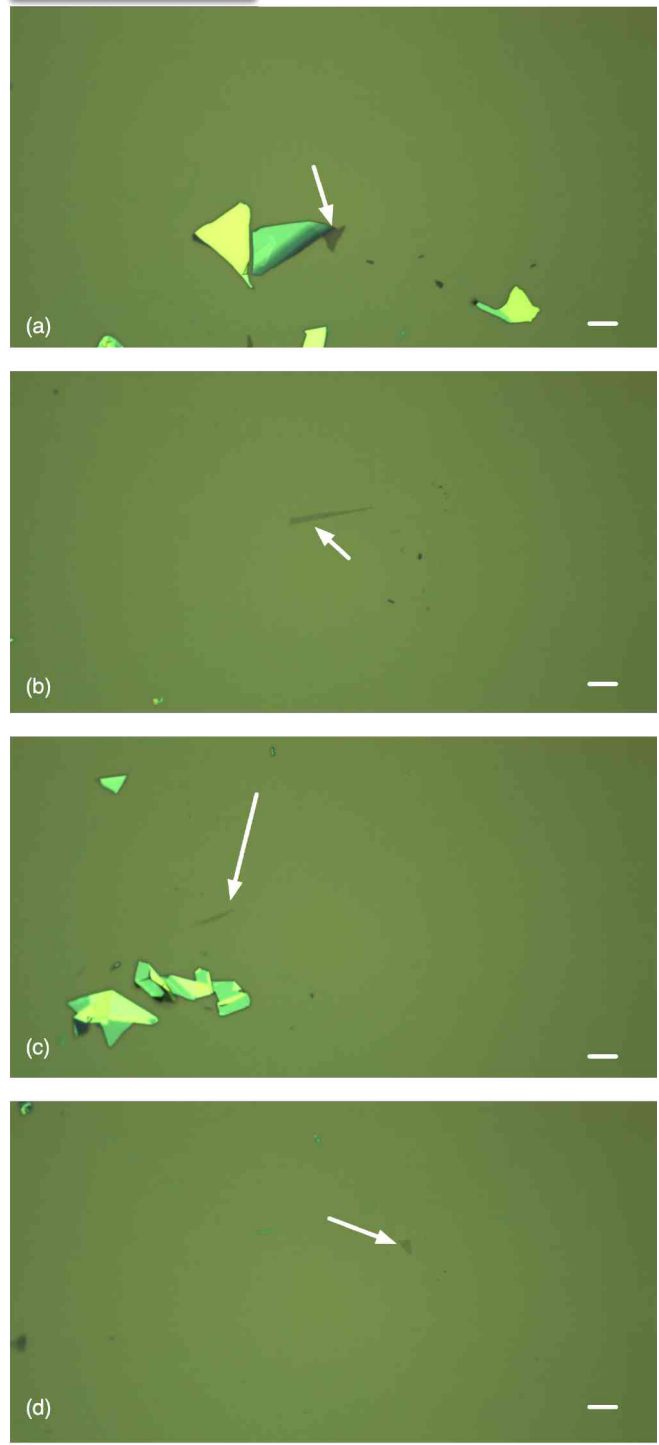

Fig. S7. Examples of false negative detection for graphene flakes. The positions of the graphene flakes are indicted by the white arrows. The scale bars correspond to 10 μm.

**Chip# 287 WTe$_2$**

| True Positive | False Positive | False Negative | True Negative |
|---|---|---|---|
| 162 | 146 | 6 | 2393 |

Precision: 0.525974026
Recall: 0.964285714

**Chip# 301 Graphene**

| True Positive | False Positive | False Negative | True Negative |
|---|---|---|---|
| 823 | 40 | 17 | 1509 |

Precision: 0.953650058
Recall: 0.979761905

Table S1. Detection performance results for WTe$_2$ and graphene.

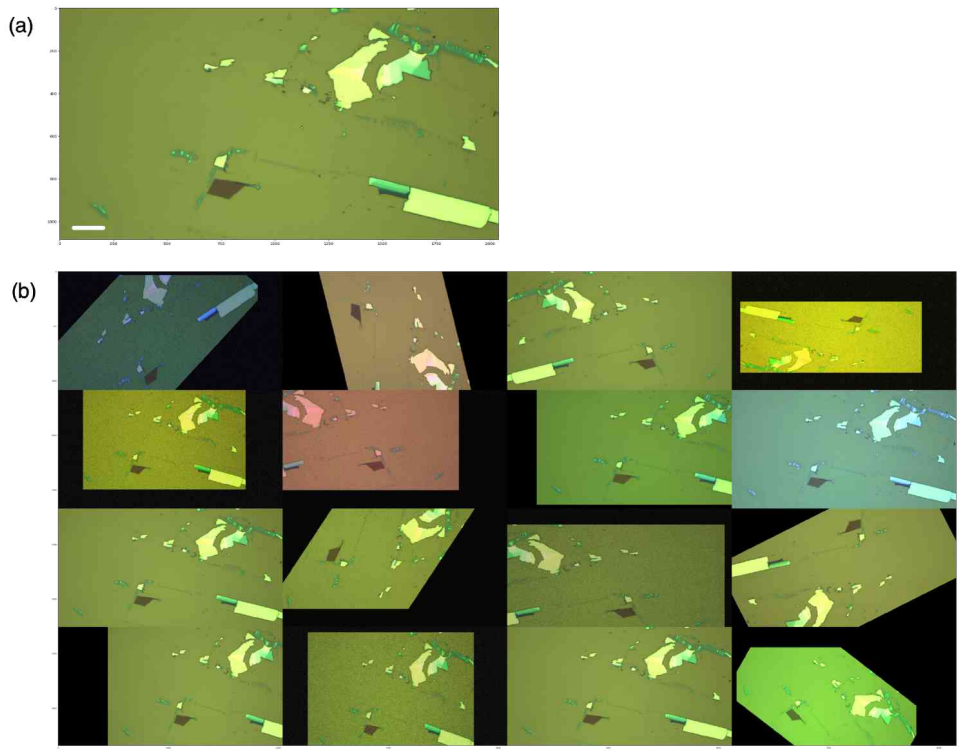

Fig. S8. Data augmentation example for WTe$_2$. (a) Input optical microscope image containing WTe$_2$ flakes and (b) augmented images. The scale bars correspond to 20 µm.

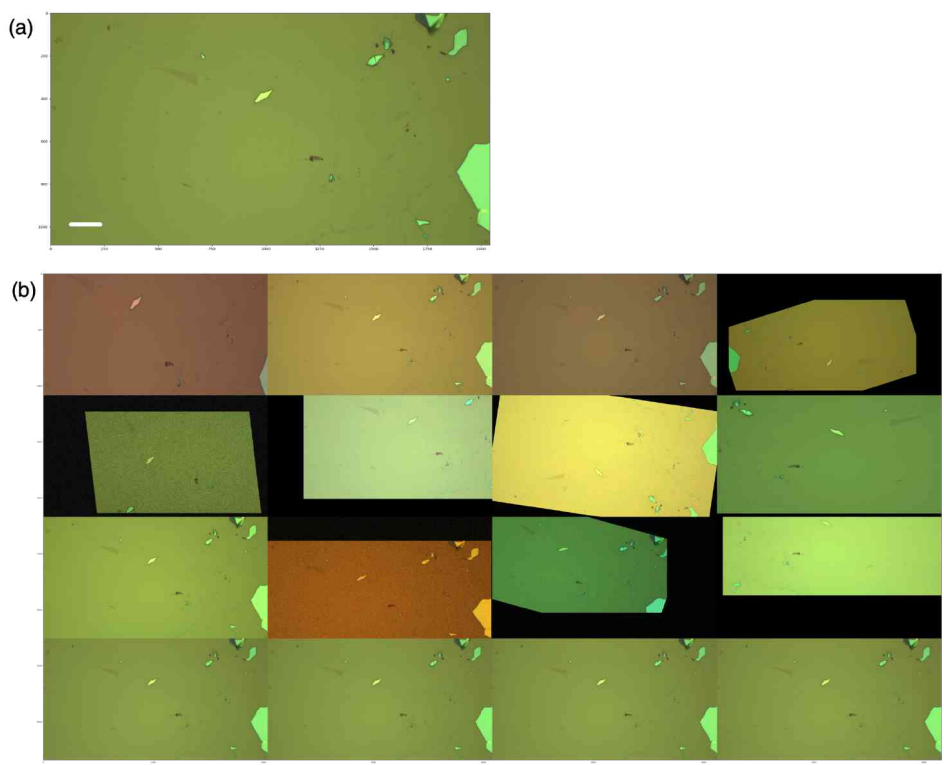

Fig. S9. Data augmentation example for graphene. (a) Input optical microscope image containing graphene flakes and (b) augmented images. The scale bars correspond to 20 μm.

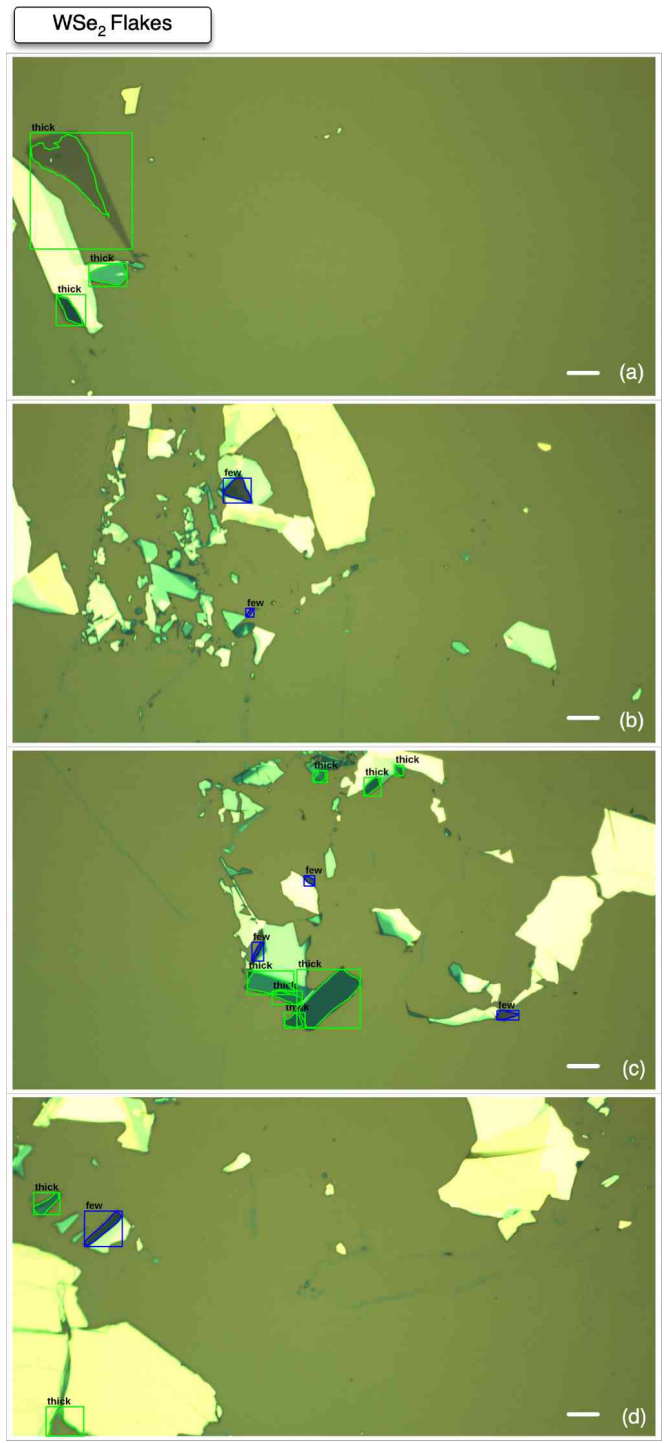

Fig. S10. Detection of $WSe_2$ flakes by deep learning inference. The scale bars correspond to 10μm.

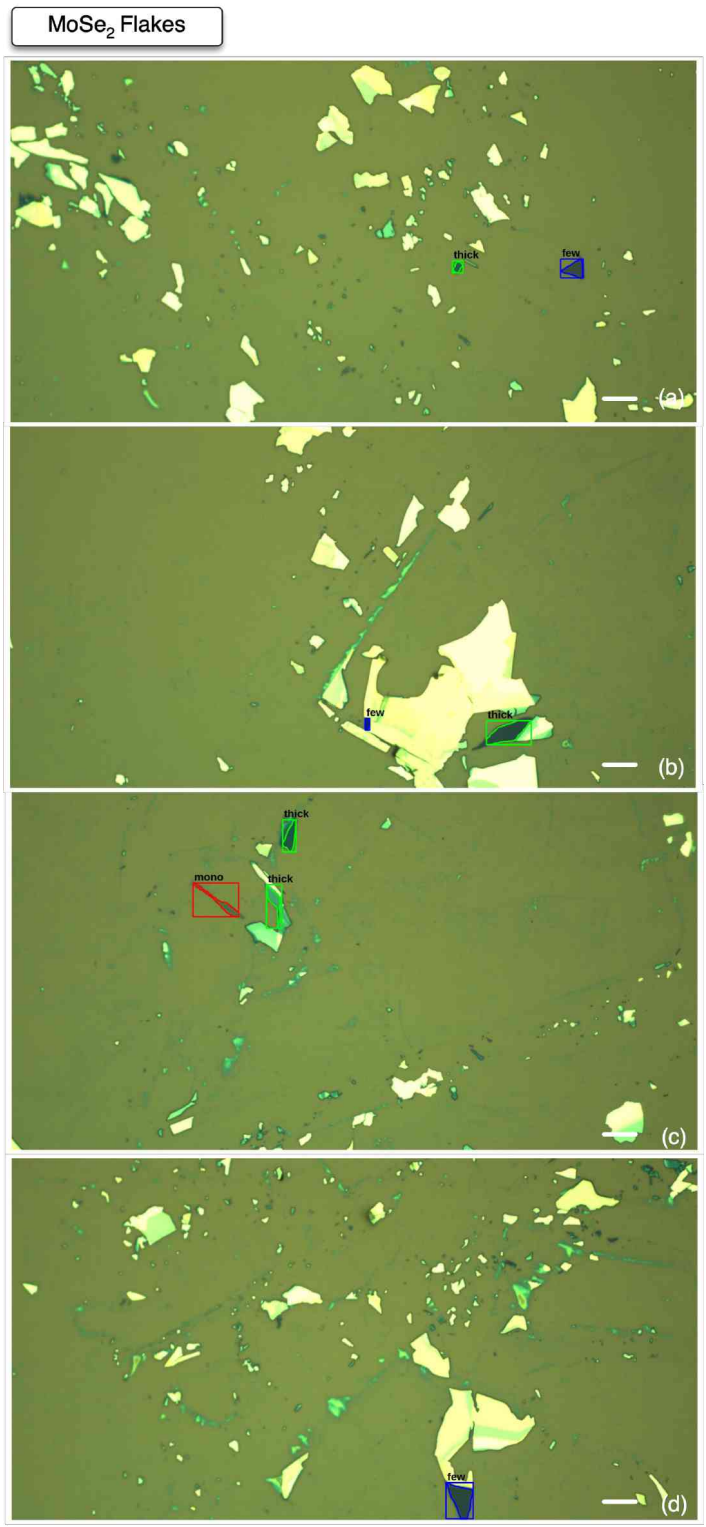

Fig. S11. Detection of MoSe$_2$ flakes by deep learning inference. The scale bars correspond to 10µm.

**Robustness against variations in magnification**

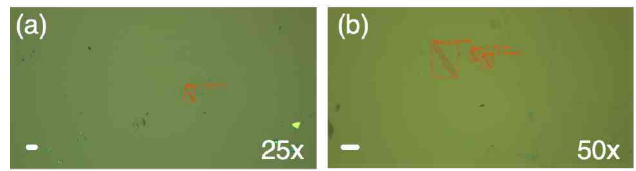

Fig. S12. Dependence of inference on changes in magnification. (a) and (b) Optical microscope images and inference results for exfoliated graphene on SiO$_2$/Si with objective lens magnifications of (a) 25× and (b) 50×. Note that monolayer graphene flakes were detected at both magnifications. The scale bars correspond to 10 μm.

**Fabrication of van der Waals heterostructures using 2D crystals detected by deep learning**

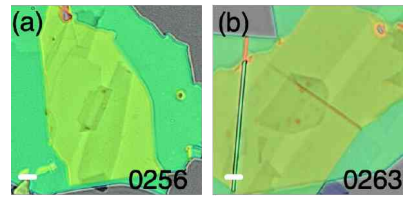

Fig. S13. Optical microscope images of van der Waals heterostructures fabricated using WTe$_2$ crystals detected by deep learning. (a) Trilayer WTe$_2$ in contact with two graphene flakes. (b) Five-layer WTe$_2$ in contact with graphene flakes. Both devices were encapsulated between hBN flakes. The scale bars correspond to 4 μm.